\documentclass[aps,pra]{revtex4}
\usepackage{graphicx}
\begin{document}
\title{High-sensitivity imaging with multi-mode twin beams}
\author{E.~Brambilla, L.~Caspani, O. Jedrkiewicz, L.~A.~Lugiato and A.~Gatti}
\address{INFM-CNR-CNISM, Dipartimento di Scienze Fisiche e Matematiche,
Universit\`a dell'Insubria, Via Valleggio 11, 22100 Como, Italy}
\begin{abstract}
Twin entangled beams produced by single-pass parametric down-conversion (PDC)  offer the opportunity 
to detect weak amount of absorption with an improved sensitivity with respect to standard techniques
which make use of classical light sources. We propose a differential measurement scheme which exploits
the spatial quantum correlation of type II PDC to image a weak amplitude object with a 
sensitivity beyond the standard quantum limit imposed by shot-noise.
\end{abstract}

\centerline{Version \today}
\maketitle

\section{Introduction}
The field of Quantum Imaging aims to exploit the quantum nature of light and the natural parallelism of optical signals to devise novel techniques for optical imaging and for parallel information processing at the quantum level.
To cite a few examples, multi-mode quantum correlations in macroscopic twin beams have been used to measure small laser beam displacements beyond the Rayleigh limit \cite{bib1} and  for the noiseless amplification of optical images \cite{bib2}. Parallely, a number of new applications that use multi-mode twin beams in the low gain regime have been proposed, such as e.g. quantum lithography \cite{bib3}, entangled photon microscopy \cite{bib4}, and dispersion canceled quantum optical coherence tomography \cite{bib5}.
The state of the art in this field can be appreciated by reading \cite{bib6}.
An old review is given by \cite{bib7}, while a new review is in press \cite{bib8}. 

In this paper we focus our attention on a specific application, namely the possibility of improving the sensibility in the measurements of very weak images, i.e. the intensity distribution transmitted by objects with a small absorption coefficient.
The detection of a weak amount of absorption (but not of its spatial distribution) with a sensitivity beyond the standard quantum limit (SQL) was demonstrated in the past by using single-mode twin beams produced by cw optical parametric oscillators (OPOs) \cite{bib9,bib10,bib11,bib12}. 
In these experiments the sub-shot-noise intensity correlation of the twin beams was exploited in order to perform differential intensity measurements of a very slight amount of absorption.
For example in \cite{bib12} this technique was used in order to measure a slight amount of absorption from an electro-optical amplitude modulator, achieving 7dB of noise reduction with respect to the SQL of a
classical differential scheme. In \cite{bib11} a spectroscopic measurement of a two-photon transition was implemented with a noise reduction of 1.9 dB. 

However, single-mode twin beams cannot be exploited to retrieve information on the spatial distribution of the transmitted field since the correlation vanishes as soon as one detects small portions of the two
beams instead of the whole beams. Being interested in measuring an image, we are forced to consider a multi-mode source which is able to display quantum correlation also in the spatial domain. Recently, our group demonstrated the existence of such kind of correlations in the high gain regime of single-pass parametric down conversion (PDC) \cite{bib13,bib7,bib14,bib15}. 
In the experiment \cite{bib14,bib15} the sub-shot noise correlation between symmetrical points of the PDC far field was observed and was interpreted as a manifestation at the macroscopic level (i.e. in the large photon number regime) of the transverse momentum conservation of photons. 
The aim of this paper is to show that this spatial twin beam effect existing over the several phase conjugate signal and idler mode pairs offers the opportunity to retrieve the full 2D spatial distribution of a weak object in parallel, with an improved signal-to-noise ratio with respect to standard techniques. This opportunity is interesting, for example in the case of biological samples or whenever there is the need for illuminating the object at low intensities.

The paper is organized in the following manner.
In Sec.\ref{sec2} we briefly review the principle of operation of differential measurements used to detect faint amounts of absorption, showing with a simple two-mode model that the use of a source displaying quantum correlations can provide a higher sensitivity than a classical source. In Sec.\ref{sec3} we propose an imaging scheme that can be used to detect the spatial distribution of amplitude objects with a PDC source. The model equations 
that describe the multi-mode PDC process are illustrated in Sec.\ref{sec4}.
In the last sections of the paper (Sec.\ref{sec4b} to Sec.\ref{sec6}) we investigate on the robustness of the spatial quantum correlation necessary to implement the high sensitivity imaging technique, considering experimental imperfections that are difficult to avoid in a real experiment. In particular in Sec. \ref{sec6} we present the results obtained through a numerical stochastic model that simulates the imaging experiment with realistic parameters.    

\section{Description of the differential detection scheme}
\label{sec2}
\begin{figure}[ht]
\includegraphics[width=10cm]{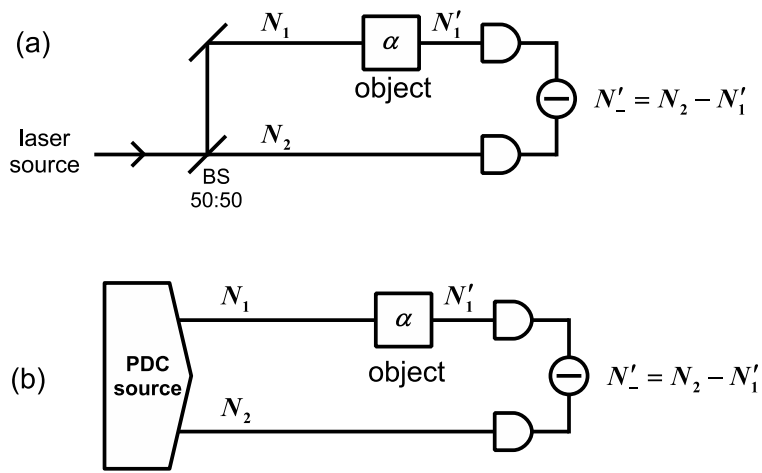}
\caption{Scheme for the detection of a weak object through a differential measurement. In (a) the test and the reference beams are obtained by splitting a laser beam with a symmetric beam splitter; in (b) the signal and idler beams of a PDC source are used.}
\label{scheme}
\end{figure}

A common procedure to detect a weak object makes use of a differential measurement. This technique is illustrated schematically in Fig.~\ref{scheme}. In its classical version [Fig.~\ref{scheme}(a)] a laser beam (not necessarily shot noise limited) is separated with a 50/50 beam splitter into two classical ``twin'' beams, a ``test'' beam which illuminates the object under examination, and a ``reference'' beam which does not interact with the object. In the ideal case in which the two beams are perfectly balanced, they can be considered as two classical copies one of each other, meaning that their intensity fluctuations display the strongest level of correlation allowed by a classical source (corresponding to the shot noise level of the incident beams). The subtraction of the detected intensities allows to eliminate the classical excess noise contained in the source field and to retrieve information about the object with a far better signal-to-noise ratio (SNR) than through direct illumination. However, when a classical source is used, this technique is limited intrinsically by the level of shot noise.

We shall now illustrate how, under appropriate conditions, quantum correlations allow to achieve a better SNR of the object than classical correlations. In the quantum scheme [see Fig.~\ref{scheme}(b)], the reference and the test beams are replaced by the signal and idler fields generated by a PDC source, for example the single-mode twin beams generated by an OPO above threshold. 

Let us indicate with $a_1$ and $a_2$ the field operators of the two beams at the plane before the object, 
and with $N_j=a_j^{\dagger}a_j$, $j=1,2$, the corresponding photon fluxes.
The fields $a_1$ and $a_2$ obey the usual canonical commutation relations
\begin{equation}
\label{commutation}
[a_i,a_j^{\dagger}]=\delta_{i,j}\;,\;\;\;\;(i,j=1,2).
\end{equation} 
The degree of correlation of the two beams is determined by the ratio between the variance of $N_-=N_2-N_1$ and the corresponding level of shot-noise, i.e.  
\begin{equation}
\label{noisered}
\sigma=\frac{\langle \delta N_-^2\rangle}{\langle N_1\rangle+\langle N_2\rangle}\;,
\end{equation}
It is equal to unity in the case of the classical copies obtained with a well balanced 50/50 beam splitter, while it can go well below unity when the PDC source is employed (it vanishes in the ideal limit of perfect quantum intensity correlations). We assume that the two beams are perfectly balanced and have symmetric statistical properties, so that in particular
\begin{equation}
\label{symmetry}
\langle N_1 \rangle=\langle N_2\rangle\;,\hspace{1cm} 
\langle \delta N_1^2 \rangle=\langle \delta N_2^2\rangle\;.
\end{equation}
Indicating with $t_{\rm obj}$ the transmission coefficient of the object, 
the field in the test arm undergoes the unitary transformation 
\begin{equation}
\label{objinputoutput}
a_1'=t_{\rm obj}\,a_1+i\sqrt{1-t_{\rm obj}^2}\,v\;,
\end{equation}
where $v$ denotes here a vacuum field operator. The measured quantity is the difference of the photon fluxes collected on the two detectors after the test beam has passed through the object, associated with the operator $N'_-=a_2^{\dagger}a_2-a_1^{\dagger}\,'a_1'$.
Recalling the condition $\langle N_1 \rangle=\langle N_2\rangle$, one immediately sees that
the mean value of $N'_-$ is proportional to the object absorption coefficient $\alpha=1-|t_{\rm obj}|^2$:
\begin{equation}
\langle N'_-\rangle=\langle a_2^{\dagger}a_2 \rangle-|t_{\rm obj}|^2\langle a_1^{\dagger}a_1\rangle=\langle N_1\rangle \alpha
\end{equation}
The noise of the measurement is determined by the variance of $N_-'$. 
By using relations (\ref{commutation}) and (\ref{objinputoutput}) together with the symmetry conditions
(\ref{symmetry}), we can express $\langle \delta N_-'^2\rangle$ in terms of the unprimed quantities $\langle N_1\rangle$, $\langle \delta N_1^2 \rangle$ and $\langle \delta N_-^2\rangle$, which depend on the photon statistics and correlation of the two beams in absence of the object:
\begin{equation}
\langle \delta N_-'^2\rangle=
\alpha^2\langle \delta N_1^2\rangle
+(1-\alpha)\langle \delta N_-^2\rangle+\alpha(1-\alpha)\langle N_1 \rangle\;.
\label{vmeno}
\end{equation}
Using (\ref{vmeno}) together with definition (\ref{noisered}) we obtain the SNR of the measurement
\begin{equation}
\label{SNRpdc}
{\rm SNR}_{\sigma}\equiv\frac{\langle N_-'\rangle}{\sqrt{\langle \delta N_-'^2\rangle}}
=\frac{\alpha\sqrt{\langle N_1\rangle}}
{\sqrt{\alpha^2 E_n + 2\sigma(1-\alpha)+\alpha}}
\end{equation}
where we have introduced the quantity
\begin{equation}
\label{excess}
E_n=\frac{\langle \delta N_1^2\rangle-\langle N_1\rangle}{\langle N_1\rangle}  
\end{equation}
which characterizes the signal (idler) noise in excess with respect to that of a coherent beam with the same intensity (it is the so called Mandel Q-factor, used when dealing with sub-poissonian statistics for which $Q$ takes negative values). 

In the classical scheme of Fig.~\ref{scheme}(a), provided the two beams are perfectly balanced and satisfy conditions (\ref{symmetry}), it can be verified that $\sigma=1$ (i.e. $\langle \delta N_-^2 \rangle=2\langle N_1 \rangle$) whatever the amount of excess noise. Thus we find
\begin{equation}
\label{SNRclass}
{\rm SNR}_{class}=
\frac{\alpha\sqrt{\langle N_1\rangle}}
{\sqrt{\alpha^2 E_n + 2-\alpha}}\;.
\end{equation} 
The standard quantum limit is obtained from this expression by considering the case of a coherent beam, for which $E_n=0$ and
\begin{equation}
\label{SQL}
{\rm SNR}_{SQL}=\frac{\alpha\sqrt{\langle N_1 \rangle}}{\sqrt{2-\alpha}}\;.
\end{equation}
Noting that for a weak object the condition $E_n<<1/\alpha^2$ is easy to achieve, the SNR obtained
with the differential measurement (\ref{SNRclass}) usually approaches the SQL value (\ref{SQL}) even if the classical source is not shot-noise limited but presents some excess noise. 
%This later feature makes the differential measurement technique advantageous for the detection of weak amount of %absorption with respect to other techniques which do not make use of a reference beam.
 
It is worth noting that the SQL could also be defined by considering the SNR of a direct measurement scheme, which makes use of a single coherent beam with the same mean photon number $\langle N_1 \rangle$: this would give ${\rm SNR}_{SQL}=\alpha\sqrt{\langle N_1 \rangle}$, an improvement by a factor $\sim\sqrt{2}$ with respect to the SQL of the differential measurement (\ref{SQL}). However in the following we will use definition (\ref{SQL}) since 
the detection of weak amount absorption is most commonly obtained through the differential measurement technique that allows to approach the SQL much more easily than direct measurements.

Let us now investigate the conditions under which the SQL can be beaten by using photon number correlated beams. To this end we introduce the ratio
\begin{equation}
\label{improvement}
R\equiv\frac{{\rm SNR}_{\sigma}}{{\rm SNR}_{SQL}}=\sqrt{\frac{2-\alpha}{\alpha^2 E_n + 2\sigma(1-\alpha)+\alpha}}
\end{equation}
which measures the improvement of the SNR achieved by using the PDC source with respect to the SQL. 
In the limit of a weak object we have
\begin{equation}
\label{weaklimit}
R \approx\frac{1}{\sqrt{\sigma}}\hspace{1.5cm}{\rm for}\;\alpha<<1\,,\;\alpha^2 E_n\ll 2\sigma
\end{equation}
Thus, provided the excess noise is not too large, the SNR is improved by a factor $\sim 1/\sqrt{\sigma}$ with respect to the SQL.  More generally, it is easily seen that $R>1$ provided that
\begin{equation}
\label{sigmamax}
\sigma<\sigma_{MAX}=1-\frac{\alpha^2 E_n}{2(1-\alpha)}
\end{equation}
$\sigma_{MAX}$ represents the maximum value of $\sigma$ for which we have an improvement in the SNR with respect to the SQL. Clearly we have always $\sigma_{MAX}<1$ (since $E_n\geq1$ and $0<\alpha<1$). This is an expected result: the twin beams of the PDC source need to be correlated below the shot noise level (i.e. $\sigma<1$) in order to beat the classical source configuration. It can be seen that condition (\ref{sigmamax}) is easily verified when $\alpha\rightarrow 0$ since $\sigma_{MAX}$ is close to unity in this limit. However the goal is to obtain a substantial improvement of the SNR and Eq.~(\ref{weaklimit}) shows that the condition on $\sigma$ is rather severe.

\section{The high-sensitivity imaging scheme}
\label{sec3}
\begin{figure}[ht]
\includegraphics[width=12cm]{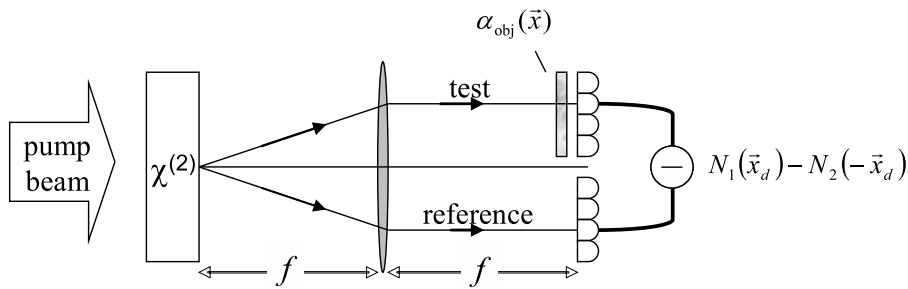}
\caption{High sensitivity imaging scheme that exploits sub-shot noise correlations between symmetrical angular
directions of the PDC field. The far field is observed in the focal plane of a lens with an $f$-$f$ imaging system. The object absorption coefficient $\alpha(\vec{x})$ is retrieved by subtracting the intensities measured from symmetrical pixels of the two arrays of detectors.}
\label{imagingscheme}
\end{figure}

The previous treatment did not consider the spatial aspects of the detection process.
Since we are interested in developing an imaging system, parallel multi-mode operation is required.
We focus here on the nonclassical source case, being interested in determining the conditions where the SQL can be beaten.
The process of spontaneous parametric down-conversion is naturally broadband in the spatial as well as in the temporal domains and is therefore well suited for this task. In particular, we can exploit the sub-shot noise photon-number correlation arising between pairs of symmetrical regions of the far field, theoretically predicted in \cite{bib16,bib13} and experimentally demonstrated in \cite{bib14,bib15}. Close to the degenerate frequency, photon pairs are indeed emitted almost symmetrically with respect to the pump axis because of the conservation of the transverse momentum. The finite waist $w_p$ of the pump beam introduces a spread in the relative angular directions of the twin photons on the order of the pump angular bandwidth $\propto\lambda/w_p$, $\lambda$ denoting the
signal/idler wavelength. 
Assuming the far field is observed in the focal plane of a lens with focal length $f$ (see Fig.~\ref{imagingscheme}) this angular spread corresponds to the coherence length of the field in this plane $x_{coh}\approx \lambda f/w_p$.
In \cite{bib16} it was shown that the fluctuations in the photon number difference detected in two symmetrical regions go well below shot-noise (ideally to zero when losses are negligible), provided the detection areas are large compared to $x_{coh}$. 

In accordance with this result, we consider the imaging scheme illustrated in Fig.~\ref{imagingscheme}. The pump field which illuminates the $\chi^{(2)}$ nonlinear crystal is a coherent pump pulse with a large beam waist $w_p$.
The object, characterized by an absorption coefficient $\alpha(\vec{x})$ depending on the transverse plane position  $\vec{x}$, is placed in the far field of the source. The lens shown in the figure is used to perform the Fourier transformation of the fields from the source into its second focal plane, which is taken as the detection plane. Both the signal (test) and the idler (reference) photons are collected by two arrays of detectors, typically the pixels of a high quantum efficiency CCD camera \cite{bib17}. The detectors are either placed immediately after the object plane as in the figure, or more realistically, a telescopic system is used to image the object plane into the detection plane. The far field coherence length $x_{coh}\approx \lambda f/w_p$ is assumed to be on the same order of magnitude of the pixel detector size or smaller, in order to fulfill the sub-shot noise correlation condition and compatibly with the requirement of large photon number operation. In this way, the fluctuations of the reference and the test beams display strong spatial ({\em point to point}) correlation for each symmetrical pair of detectors of the CCD array. Sub-shot noise correlation for each of these pairs of detection channels is the goal to achieve in order to beat the SQL in a differential measurement. 

\section{The numerical model}
\label{sec4}
With the aim of simulating the high-sensitivity imaging experiment proposed in the previous section, we developed a fully 3D numerical model which takes into account realistic features of the PDC process, such as the phase-matching conditions inside the crystal as well as the finite size of the pump beam both in the temporal and the spatial domains.  The model is based on a set of equations describing the propagation of the slowly varying envelope operators of the
signal and idler fields, which we denote by $a_1(z,\vec{x},t)$ and $a_2(z,\vec{x},t)$, and
satisfy boson-like commutation rules at equal $z$, i.e.
$[a_i(z,\vec{x},t),a_j^{\dagger}(z,\vec{x},t)]=\delta_{i,j}\delta(\vec{x}-\vec{x}')\delta(t-t')$, $(i,j=1,2)$. 
As we are considering the parametric regime, it is assumed that the pump field $A_{p}(\vec{x},t)$
remains almost undepleted and can be treated as a known classical field. For definiteness, 
we shall assume it has a Gaussian profile with a plane-wave front at $z=0$, with a beam waist $w_{p}$ and
duration $\tau_{p}$:
\begin{equation}
\label{pump}
A_0(z=0,\vec{x},t)=(2\pi)^{-3/2}A_p e^{-(x^2+y^2)/w_p^2}e^{-t^2/\tau_{p}^2}\;.
\end{equation}
The propagation equation for the signal field $a_1$ can then be written in Fourier space 
as \citet{bib16,bib18}:
\begin{eqnarray}
\label{waveq}
\frac{\partial}{\partial z} a_1(z,\vec{q},\Omega) &=&
    i\left[k_{1z}(\vec{q},\Omega)-k_1\right]a_1(z,\vec{q},\Omega)\\
    &+&\frac{g}{l_c} e^{-i\Delta_0 z}
    \int d\vec{q}~' \int d\Omega'
         \alpha_p(z,\vec{q}-\vec{q}~',\Omega-\Omega')a_2^{\dag}(z,-\vec{q}~',-\Omega')
    \nonumber\;,
\end{eqnarray}
the equation for the idler wave $a_2$ can be obtained by exchanging
the subscript $1\leftrightarrow 2$. The operators $a_j(z,\vec{q},\Omega)$ denotes the
Fourier transformation of the fields $a_j(z,\vec{x},t)$:
\begin{equation}
a_j(z,\vec{q},\Omega)=
     \int\frac{d\vec{x}}{2\pi}\int\frac{dt}{\sqrt{2\pi}}
                 a_j(z,\vec{x},t)e^{-i\vec{q}\cdot\vec{x}+i\Omega t}
     \hspace{1cm}(j=1,2)\;.
\end{equation}
where $\Omega$ denotes the temporal frequency relative to the carrier frequency $\omega_j$ of field $j$,
$\vec{q}$ the transverse wave vector component; $k_{jz}(\vec{q},\Omega)=\sqrt{k_j(\vec{q},\Omega)-q^2}$ is the corresponding longitudinal component along the propagation direction $z$.
The wave-number function  $k_j(\vec{q},\Omega)$ determines the linear propagation properties of field $j$; in particular its linear and quadratic dependence on $\Omega$ is responsible of temporal walk-off and second-order temporal dispersion, while its linear and quadratic dependence on $\vec{q}$ leads to spatial walk-off and diffraction. 
The parametric gain $g$ is a dimensionless constant proportional to the strength of the nonlinear interaction, to the
crystal length $l_c$, and to the peak value of the pump field $A_p$; its value determines the number of photons that are generated in the down-conversion process in mode pairs that are well-phase-matched.
$\Delta_0=k_1+k_2-k_p$ indicates the collinear phase-mismatch parameter of the three carrier waves. The normalized function $\alpha_p(\vec{q},\Omega)=1/(\pi^{3/2}\delta q_p^2\delta \omega_p) e^{-q^2/\delta q_P^2}e^{-\omega^2/\delta \omega_P^2}$ is the Fourier transform of $A_0(\vec{x},t)/A_p$, with $\delta q_p=2/w_p$ and $\delta \omega_p=2/\tau_{p}$
denoting the pump spatial and temporal bandwidths respectively. 

In the plane-wave pump (PWP) limit where $\delta q_p$ and $\delta \omega_p$ are much smaller than the spatial and temporal bandwidths of phase-matching, $\alpha_p(\vec{q},\Omega)\rightarrow\delta(\vec{q})\delta(\Omega)$ and the
model equations can be solved analytically.  The solution of Eqs.~(\ref{waveq})
can be written as an input-output relation in the form of a two-mode squeezing transformation:
\begin{eqnarray}
\label{inout}
a_1^{out}(\vec{q},\Omega)&=&U_1(\vec{q},\Omega)a_1^{out}(\vec{q},\Omega)
                           +V_1(\vec{q},\Omega)a_2^{\dagger in}(-\vec{q},-\Omega)\;,\nonumber\\ 
a_2^{out}(\vec{q},\Omega)&=&U_1(\vec{q},\Omega)a_1^{out}(\vec{q},\Omega)
                           +V_1(\vec{q},\Omega)a_2^{\dagger in}(-\vec{q},-\Omega)\;, 
\end{eqnarray}
The explicit expressions of the gain coefficient $U_j$ can be found in \cite{bib16,bib18}.
Relation (\ref{inout}) links only phase-conjugate modes $(\vec{q},\Omega)_1$ and $(-\vec{q},-\Omega)_2$ of the signal and idler fields. This feature reflects the perfect correlation of the transverse momenta and energies of the emitted twin photons which occurs in the PWP limit. In particular, the signal-idler spatial correlation function is found to display a delta-like peak in correspondence to opposite positions in the far field plane, as a consequence of the transverse momentum conservation rule \cite{bib13,bib16}).

As already mentioned in the previous section, the finite pump bandwidth $\delta q_p$ in the spatial domain introduces an indeterminacy in the correlation between the angular directions of the emitted twin photons. 
This indeterminacy gives rise to a spread of the signal-idler correlation function which acquires a finite width
on the order of the far field coherence length $x_{coh}\approx\lambda f/w_p$. 
Thus, as a consequence of the finite transverse size of the pump beam, perfect spatial correlations between two symmetrical ideal detectors can be achieved only if the linear size of the detectors is 
large compared to $x_{coh}$; in this ideal limit we would have $\sigma=0$. Conversely, if this condition is not fulfilled the fluctuations of $N_-$ just lie below or close to the shot-noise level, i.e. we have $0<\sigma<1$. When the finite quantum efficiency of the detectors $\eta<1$ is taken into account, the highest level of correlation that can be reached is $\sigma=1-\eta$ (see e.g. \cite{bib13}).

In the simulations we shall consider an optical setup close to that of the experiment performed in Como to measure sub-shot-noise far field correlations in the high gain regime of PDC \citet{bib14,bib15}: the nonlinear crystal is a 4\,mm long beta-barium borate (BBO) crystal cut for type II phase-matching and PDC is observed around the degenerate wavelength $\lambda=704$\,nm. The far field is measured in the focal plane of a lens of focal length $f=5$cm as shown in Fig.~\ref{imagingscheme}. 
The pixel detectors of the CCD are characterized by a 20x20 $\mu$m$^2$ square area and a high quantum efficiency $\eta$ approaching 0.9 at the signal wavelength. The effective quantum efficiency which takes into account the losses of all optical elements was evaluated to be around 0.75 \cite{bib14}. 

All the numerical results presented in the following sections have been obtained by considering a stochastic model based on the Wigner representation equivalent to Eqs.~(\ref{waveq}). All the expectation values
of the field operators moments necessary to evaluate the observables of interest
were evaluated by integrating numerically the classical-looking propagation equations of the model (formally identical to Eqs.~(\ref{waveq})) and by performing the necessary stochastic averages on the output fields. As the Wigner representation provides only symmetrically ordered operator moments, we had to apply appropriate corrections to obtain the desired ordering (see e.g. \cite{bib16,bib19} for more details).

\section{Behaviour of the signal-idler far field correlation function}
\label{sec4b}
In this section we briefly discuss some aspects of the signal-idler intensity correlation that are relevant
for the proposed imaging scheme. The spread of this function indeed determines the minimum size of the detection area for which it is possible to observe sub-shot noise correlation and sets therefore a lower limit for the imaging resolution. To be more precise, let us consider the normalized correlation function for the signal and idler intensity fluctuations, defined as
\begin{equation}
\Gamma_{12}(\vec{x}_1,\vec{x}_2)=\frac{ \langle \delta N_1(\vec{x}_1) \delta N_2(\vec{x}_2) \rangle}
                   {\sqrt{\langle \delta N_1(\vec{x}_1)^2\rangle\langle \delta N_2(\vec{x}_2)^2\rangle}}\;,
\end{equation}
where $\delta N_j(\vec{x}_j)=N_j(\vec{x}_j)-\langle N_j(\vec{x}_j) \rangle$, $(j=1,2)$ denotes the photon number fluctuation operator in position $\vec{x}_j$ of the detection plane for field $j$. 
Notice that with this normalization $|\Gamma_{12}|\le 1$ and $\Gamma_{12}=1$ represents the maximum possible amount of correlation. As discussed in the previous section, the correlation function displays a peak for $\vec{x}_2=-\vec{x}_1$ as a consequence of the $\vec{q} \leftrightarrow -\vec{q}$ momentum correlation of the emitted twin photons. 

\begin{figure}[ht]
\centering
\includegraphics[width=11.0cm]{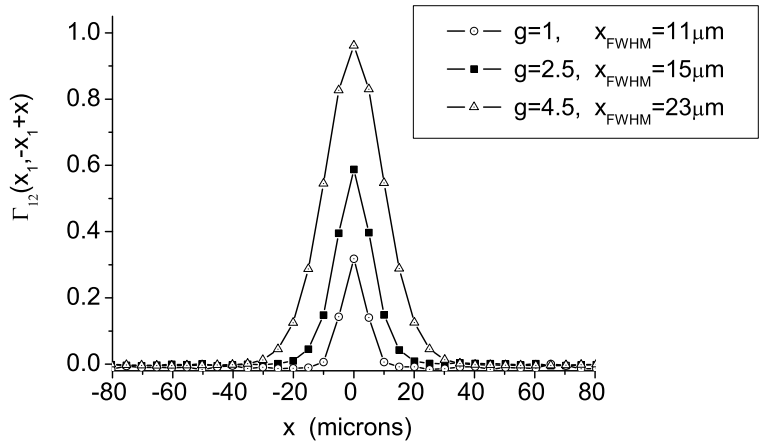}
\caption{Cross-section of the normalized signal-idler correlation function $\Gamma_{12}(\vec{x}_1,-\vec{x}_1+\vec{x})$ as evaluated from the stochastic numerical model for increasing values of $g$. The pump beam waist is $w_p=1500\mu$m, the lens focal length $f=5$cm. The spatial step size of the numerical grid is $5\mu$m.}
\label{cross}
\end{figure}
At low gains, i.e. for $g<<1$, we can use a first order perturbative expansion of the solution of Eqs.~(\ref{waveq}) in power of $g$ in order to evaluate $\Gamma_{12}(\vec{x}_1,\vec{x}_2)$. It is found \cite{bib20} that it is proportional to the square modulus of the pump Fourier transform $\alpha_{p}(\vec{q},\Omega)$  evaluated in $(\vec{x_1}+\vec{x_2})\lambda f/2\pi$. According to this result, and considering the Gaussian pump profile of Eq.~(\ref{pump}), the FWHM of the correlation function is then given by 
\begin{equation}
\label{xcoh}
x_{\rm FWHM}=\frac{\sqrt{2\ln{2}}}{\pi}\frac{\lambda f}{w_p}\;,
\end{equation}
a value corresponding to the typical far field coherence length $x_{coh}$.
With the parameter values $f=5cm$, $\lambda=0.704\mu$m and $w_p=1500\mu$m, we obtain $x_{\rm FWHM}=8.8\mu$m.

For values of $g$ on the order of unity or higher, no analytical expression of $\Gamma_{12}$ is known and we have to
resort to the numerical stochastic model described in the previous section. Fig.~\ref{cross} shows the result
of these simulations: the cross-section of $\Gamma_{12}(\vec{x}_1,-\vec{x}_1+\vec{x})$ discretized on the numerical grid is plotted as a function of $\vec{x}\equiv(x,y)$ along the $x$-axis (corresponding to the walk-off direction) for increasing values of the gain parameter $g$. 
It has been evaluated by performing ensemble averages as well as spatial averages, the latter being allowed by the translational invariance of the signal and idler fields in the considered far field region. In order to have a significant resolution, here and in the results that follows we consider a numerical grid with a spatial step size of $5\mu$m in the far field plane (according to the mapping $\vec{q}\leftrightarrow(\lambda f/2\pi)\vec{q}$ holding between the spatial frequency plane and the focal plane of the $f$-$f$ lens system illustrated in Fig.\ref{imagingscheme}).  
As can be seen from the figure legend, the smallest value $g=1$ yields a FWHM of $\sim 11\mu$m, close to the low gain limit result (\ref{xcoh}). However, by increasing the parametric gain we observe that the correlation function becomes broader and broader: its FWHM is more than doubled with respect to the low gain value as $g$ is raised to 4.5. This stronger spread occurring at high gains can be explained with the following intuitive arguments. Inside the crystal, the cascading effect which causes the exponential grow of the number of generated photon pairs is enhanced in the regions where the pump field takes its highest values. 
Thus, in a regime of very high gain, most of the photon pairs are produced where the pump field is close to its peak value, i.e close to the center of the beam. As a result the effective region of amplification inside the crystal becomes narrower than the pump beam profile (see also \cite{bib21,bib15}), and this effect produces a broadening of the signal-idler intensity correlation function in the far field.

Finally we note that the peak value $\Gamma_{12}^{max}$ of the discretized correlation function is related to the the correlation factor between symmetrical pixels, $\sigma$, and the excess noise on those pixels, $E_n$, through the relation $\Gamma_{12}^{max}=1-\sigma/(E_n+1)$; this explains the behaviour of $\Gamma_{12}^{max}$ which approaches unity as $g$ (and hence $E_n$) is increased (see Fig.\ref{cross}), although the degree of correlation (measured by $\sigma$) usually  deteriorates at high gains, as we shall show in the next section.

\section{Fragility of quantum correlation toward imperfect detection}
\label{sec5}
This section aims at identifying regimes suitable for high sensitivity image detection, which requires 
quantum correlation and a large number of photons per pixel at the same time.
The latter condition is necessary because of the unavoidable presence of detection noise. If we denote by 
$\langle \delta N_-^2 \rangle_{bkg}=\langle \delta N_1^2\rangle_{bkg}+\langle \delta N_2^2\rangle_{bkg}$
the variance of the difference of background noise from symmetrical pixels of the CCD, we need to achieve
$\langle \delta N_-^2\rangle_{bkg}/\langle N_1+N_2\rangle\ll 1$.
As it was clearly illustrated in \cite{bib14,bib15}, the task to achieve strong local sub-shot noise correlation with
a large number of detected photons is experimentally demanding. 
We wish to investigate under which conditions the SNR obtained with the PDC source is able to beat the SQL defined in Eq.~(\ref{SQL}). In particular we shall emphasize the negative role of the excess noise present in the signal and idler beams, showing that the quantum nature of the correlation becomes fragile against unavoidable imperfections in the
detection procedure. Because of this feature, the conditions of high parametric gain and high excess noise in the experiment described in \cite{bib14} would not suit our purpose. Indeed, the experimental data displayed a transition
from quantum (sub-shot noise) to classical (above shot noise) correlation as the number of PDC
photons per pixel was raised above 15-20, a regime where the intrinsic noise of the CCD is not negligible. 

It is well known that the signal and idler fields taken separately display a thermal-like statistics.  
The variance of the photon number intercepted by the pixel detectors can therefore be written in the form \cite{bib22}
\begin{equation}
\label{thermal}
\langle \delta N_2^2 \rangle=\langle N_2 \rangle +\frac{\langle N_2 \rangle^2}{M}
\end{equation}
where $M$ denotes the degeneracy factor representing the number of spatial and temporal modes collected by the
detectors. (Accurate measurements of the photon statistics in the high gain regime of PDC 
are described e.g. in \cite{bib23}.) 
According to definition  (\ref{excess}), it follows therefore that the excess noise coincides with the mean number of photons per mode: 
\begin{equation}
E_n=\frac{\langle N_2 \rangle}{M}\;.
\label{Enthermal}
\end{equation}
This quantity depends only on the parametric gain and is on the order of $\sinh^2 g$.

A large amount of excess noise has a detrimental effect on quantum correlation as one takes into account small unbalances between the test and the reference arms of the imaging system, which in real experiments can never be completely suppressed. 
Let assume for example that the signal and the idler fields are not perfectly balanced, but rather undergo different losses in the test and the reference arms, either during propagation, or in the detection process. Indicating
with $\eta_1$ and $\eta_2$ the different effective quantum efficiencies in the two detection channels, it is found that
\begin{equation}
\label{unbalanced}
\sigma=1-\bar{\eta}+\frac{(\eta_1-\eta_2)^2}{2\bar{\eta}} 
\left(E_n+\frac{1}{2}\right)\;,
\label{unbalance1}
\end{equation}
where $\bar{\eta}=(\eta_1+\eta_2)/2$. 
Thus we see from relation ($\ref{unbalance1}$) that a necessary condition to have sub-shot-noise correlation is that  $E_n\ll 2\bar{\eta}/(\eta_1-\eta_2)^2$. Because $(\eta_1-\eta_2)^2$ is usually small compared to unity, the condition can be generally fulfilled, e.g. by applying a careful compensation of the losses in the two detection  
channels \cite{bib10,bib12}.

However, when dealing with an imaging scheme as in our case, another source of unbalance which is usually more difficult to control experimentally derives from the inaccuracy in the determination of the center of symmetry of the far-field pattern in the detection plane. Typically, the distance $x_{shift}$ between the selected and the exact centre of symmetry
is on the order of one fourth the size of a CCD pixel ($20\mu$m in the experiment \cite{bib14}) and is small compared with the far field coherence length $x_{coh}$ (a few tens of microns in \cite{bib14}) but not negligible.
In the following, all the quantities of interest will refer to a pair of pixel detectors located at symmetrical positions with respect to the optical axis as the one shown in Fig.~\ref{imagingscheme}. 
In particular $N_1$ and $N_2$ will indicate the operators associated with the number of photons measured by such a pair of detectors.

\begin{figure}[ht]
\centering
\includegraphics[width=15.0cm]{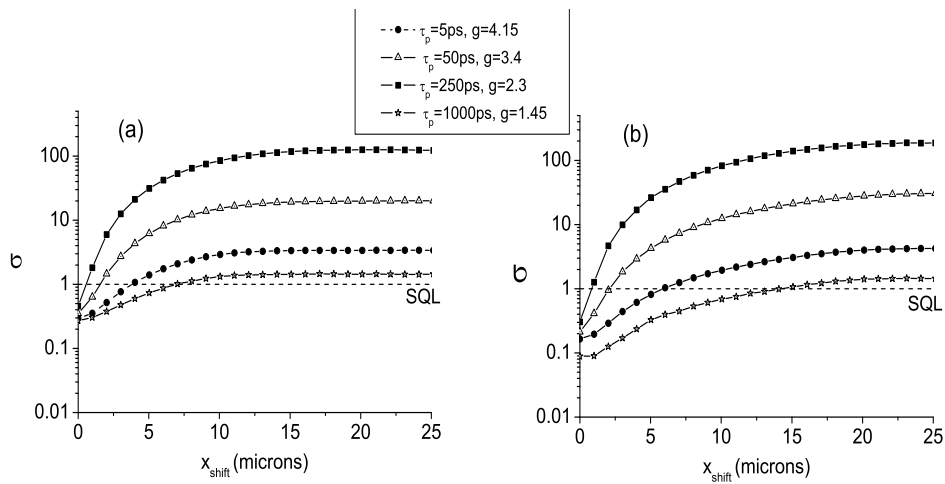}
\caption{Plot of the correlation factor $\sigma$ as a function of $x_{shift}$ for increasing values of the pump pulse duration, considering  a pixel size of $20\mu$m (a) and $40\mu$m (b). The value of the parametric gain (see legend) changes with $\tau_{p}$ so that the number of detected photons is fixed, about 3500 ph./pixel in case (a). 
The pump beam waist is $w_p=1500\mu$m, the lens focal length $f=5$cm, the detection quantum efficiency $\eta=0.9$.}
\label{sigmavsxshift}
\end{figure}
Using the numerical model illustrated in Sec.\ref{sec4}, we investigated the behaviour of the correlation factor $\sigma$ toward errors in the center of symmetry $x_{shift}$, considering different values of the parametric gain $g$
in order to vary the amount of excess noise.  
The far field measurement scheme considered in the simulation is identical to the one illustrated in Fig.~\ref{imagingscheme},  except we do not include the absorbing object, as we are presently interested only in evaluating the correlation factor.  The results are illustrated in Fig.~\ref{sigmavsxshift}. We considered decreasing values of the parametric gain $g$ in order to reduce progressively the amount of excess noise (which behaves as $\sinh^2 g$) starting from rather high values ($E_n\sim100$ for $g=4.15$) down to a small amount ($E_n\sim 0.5$ for $g=1.45$). At the same time the pump pulse duration $\tau_p$ is increased as indicated in the figure legend so that the number of photons $\langle N_1 \rangle$ detected by the pixel detectors remains unchanged, about 3500\,ph./pixel. In such conditions, the excess noise decreases as the inverse of the pump pulse duration $\tau_{pump}$. 
The two plots corresponds to two different sizes of the pixel detectors: $20\mu$m in case (a) and $40\mu$m in case (b). Referring to the $20\mu$m pixels of the CCD used in experiment \cite{bib14}, the $40\mu$m  pixel size can be obtained by performing a 2x2 pixel binning. With our choice of the lens focal length ($f=5cm$) and the pump beam waist ($w_p=1550\mu$m), the typical coherence length in the detection plane is of the same order of magnitude of the pixel size or slightly smaller. 
Except for the case $\tau_{p}=5$ps, the long duration of the pump pulse with respect to the PDC coherence time ($\tau_{coh}\sim 1$ps) did not allow to simulate PDC propagation using a unique numerical array. For this reason we divided the pump temporal profile in a sequence of 5\,ps long intervals over which the generated PDC field can be considered as nearly stationary. We performed simulations for each of these time intervals separately, summing up their independent contributions in order to evaluate the overall photon statistics.   

As can be inferred by comparing  Fig.~\ref{sigmavsxshift}(a) and (b), under the same gain/pulse duration conditions the 2x2 pixel binning [Fig.~\ref{sigmavsxshift}(b)] allows to enhance the correlation substantially 
(as long as $x_{shift}\ll x_{coh}$) since the detection area covers a larger portion of the coherence area. Another relevant issue which can be inferred from the both plots is the deterioration of the correlation that occurs at high parametric gains even for $x_{shift}=0$. This effect is related to the broadening of the signal-idler correlation function $\Gamma_{12}$ which occurs at high parametric gains as discussed in Sec.\ref{sec4b}. It becomes particularly relevant for the 5ps pump pulse with $g=4.15$ and the $20\mu$m pixel detectors (black squares in Fig.~\ref{sigmavsxshift}(a):  the correlation factor $\sigma$ goes rather close to unity, i.e. close the shot noise level, since for such a high value of $g$ the FWHM of the $\Gamma_{12}$, $x_{\rm FWHM}\sim 23\mu$m, becomes larger than the size of the pixel detectors.  On the other side, considering the $40\mu$m pixel detectors, we see that $\sigma$ approaches the optimal value $1-\eta=0.1$ as the pump pulse duration is increased and the parametric gain is decreased.  

From the same plots we see that $\sigma$ starts from its minimum value taken at $x_{shift}=0$ and increases with $x_{shift}$ until it saturates to the value $\sigma_{sat}=1+E_n$ as $x_{shift}$ becomes larger than $x_{coh}$. It can be easily verified that the saturation value $\sigma_{sat}$ corresponds to the limit in which the photon number fluctuations revealed on the two detectors becomes completely uncorrelated, as the error in the symmetry center becomes larger than the width of signal-idler correlation function (which can be identified with $x_{coh}$). 
Considering the worst case $\tau_{p}=5$ps, the large parametric gain makes the excess noise extremely high ($E_n\sim 100$). 
As a consequence, we see that the slightest inaccuracy in the position of the symmetry center leads to a transition of the degree of correlation from sub-shot noise to strongly above shot noise (black square in Fig.~\ref{sigmavsxshift}). As the parametric gain is decreased $\sigma$ saturates slightly faster (since $x_{coh}$ diminishes) but to lower and lower values which tend to approach the shot noise level as $E_{n}$ goes to zero (see stars of the $1000$ps pump pulse case). As $M\propto\tau_{p}$ becomes larger than $\langle N_1 \rangle$ the photon statistics on the two detectors becomes indeed nearly  Poissonian, so that $\langle \delta N_1^2 \rangle\approx\langle N_1 \rangle$ (see Eq.~(\ref{thermal})) and $\sigma$ is at worse on the order of unity, the value corresponding to the standard quantum limit. 

If we approximate the curves of Fig.~\ref{sigmavsxshift} with straight lines that go from the minimum value $\sim 1-\eta$ (at $x_{shift}=0$) to the saturation value $\sigma_{sat}=1+E_n$ (at $x_{shift}=x_{coh}$), we can
obtain a rough estimate of the error shift value for which the transition 
from sub-shot-noise to above shot-noise correlations occurs, i.e. 
\begin{equation}
\label{shifted2}
\sigma>1\longleftrightarrow\frac{x_{shift}}{x_{coh}}>\frac{1}{1+E_n/\eta}
\end{equation}
A more precise analytical derivation of the dependence of $\sigma$ on $x_{shift}$ will be given in \cite{bib25};
however we can infer from the behavior illustrated in Fig.~\ref{sigmavsxshift} and from expression (\ref{shifted2}) that the precision in the determination of the symmetry center  becomes an extremely relevant issue from the experimental point of view whenever the excess noise is a large quantity, i.e. for short pump pulses and high parametric gains.
Moreover, relation (\ref{shifted2}) does not takes the broadening of the correlation function occurring with increasing $g$ and therefore the fragility of the quantum correlation is underestimated.
Clearly, a way to compensate for this negative effect would be to increase the detection area.
On the contrary, when the excess noise is lowered by decreasing the parametric gain, the quantum correlation becomes more robust. 
  
\begin{figure}[ht]
\centering
\includegraphics[width=8.0cm]{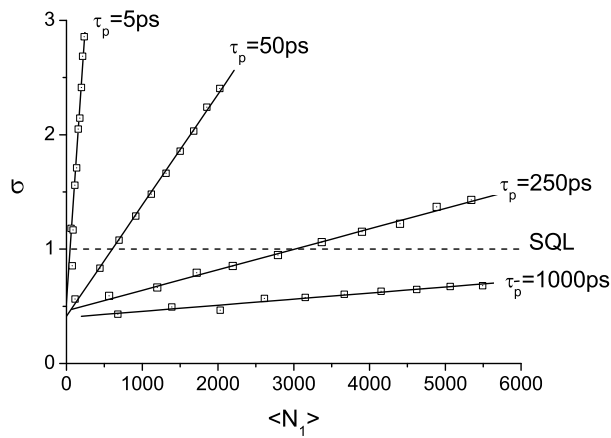}
\caption{Plot of the correlation factor $\sigma$ as a function of $\langle N_1 \rangle$ for different values of the pump pulse duration $\tau_{p}$ (see legend). The error in the symmetry center is $x_{shift}=4\,\mu$m. $\langle N_1\rangle$ is increased by varying the parametric gain $g$. The pixel detector size is $20\mu$m. The other parameters are the same as in Fig.~\ref{sigmavsxshift}.}
\label{longpulse}
\end{figure}
The previous results clearly suggests that sub-shot noise correlation can be maintained at higher photon number values
if the parametric gain is reduced and the pump pulse duration is increased, typically in the nanosecond range.
The simulation shown in Fig.~\ref{longpulse} confirms this hypothesis: it plots $\sigma$ as function of the mean photon number per pixel $\langle N_1 \rangle$ considering a fixed error in the determination of the symmetry center, $x_{shift}=4$ $\mu$m, and for increasing values of the pump pulse duration, starting from 5ps up to 1ns. 
Each point of the curves is obtained for different values of the parametric gain $g$ in the 
the range between 1.2 and 3.5, while the solid lines correspond to a linear fit of the numerical data. 
As a general behaviour, we see from the figure that $\sigma$ increases linearly with $\langle N_1 \rangle$ with a slope which decreases as the pump pulse duration is increased. We verified that the angular coefficient of the linear fits (solid lines) scales as the inverse of $M\sim\tau_{p}/\tau_{coh}$.  
For the 1\,ns pulse one can therefore reach sub-shot noise correlation at much higher photon number values
than for shorter pulses.

\section{Simulation of the high-sensitivity imaging experiment}
\label{sec6}
From the previous results we have seen that in the presence of excess noise a careful balance of the two detection 
channels and a precise determination of the symmetry center are necessary in order to approach the (ideal) 
minimum value $\sigma=1-\eta$. 
On the other hand, we see both from Eq.~(\ref{unbalanced}) and relation (\ref{shifted2}) that this requirement becomes less stringent as the excess noise is reduced by lowering the parametric gain, so that $M$ becomes on the same order of magnitude of  $\langle N_1 \rangle$ or larger. 
We can take advantage of this feature by considering long pump pulses, typically in the nanoseconds range, so that 
$\tau_p \gg \tau_{coh}$ and a large number of temporal modes are amplified at a low parametric gain, keeping large the total number of photons collected by the pixel detectors.

\begin{figure}[ht]
\centering
\includegraphics[width=15cm]{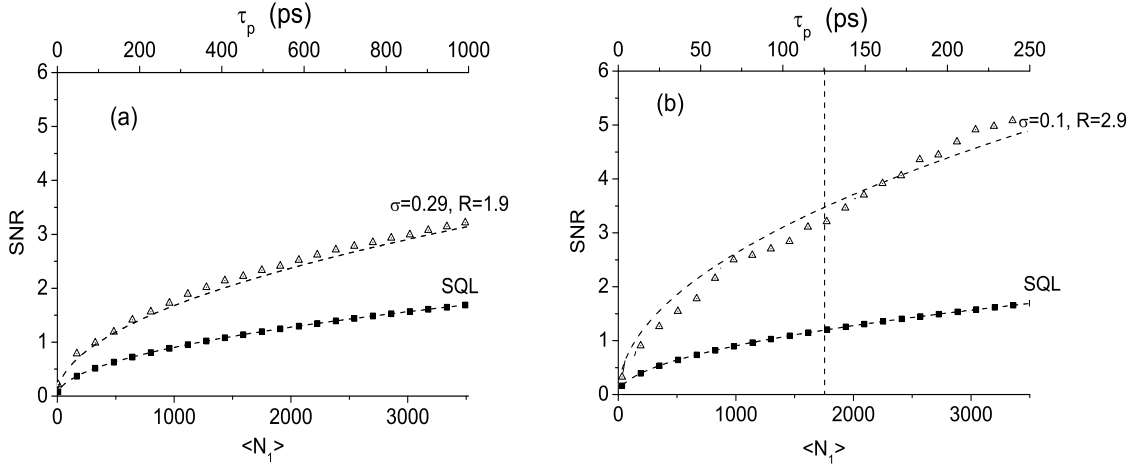}
\caption{Evaluation of the SNR obtained by increasing the pump pulse duration for a fixed parametric gain $g=1.45$. The object absorption coefficient is $\alpha=0.04$. In (a) the pixel size is $20\mu$m, in (b) a 2x2 binning is performed giving a pixel size of $40\mu$m. The lower horizontal scale gives the number of detected photons per pixel, the upper scale the corresponding pump pulse duration. The evaluated degree of correlation $\sigma$ and the SNR improvement $R$ are indicated. The test and reference arms are assumed to be perfectly balanced (in particular $x_{shift}=0$). The other parameters are the same as in Fig.~\ref{sigmavsxshift}.}
\label{SNRsim}
\end{figure}
\begin{figure}[ht]
\centering
\includegraphics[width=8.0cm]{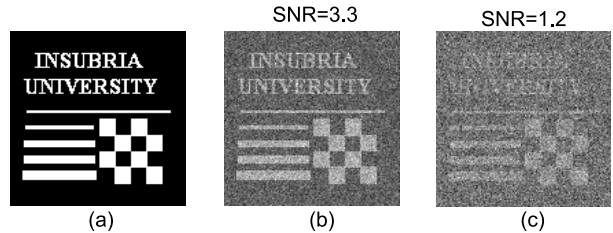}
\caption{Simulation of the retrieval of a weak object (a) through a differential measurement with a 125ps pump pulse and $\sim 1750$\,ph./pixel. The indicated SNR values corresponds to those with the PDC source (b) and with the classical source (c) in the simulation of Fig.~\ref{SNRsim}(b) with the $40\mu$m pixel size (see dotted vertical line in this figure).} 
\label{chart}
\end{figure}

We now illustrate the numerical simulation of a high-sensitivity experiment performed considering the measurement scheme of Fig.~\ref{imagingscheme}. Let us first consider the case in which test and reference detection channels are perfectly balanced (in particular $x_{shift}=0$). The object we considered took the form of a weakly absorbing mask on the numerical grid with a constant absorption coefficient $\alpha=0.04$, as the one shown in Fig.~\ref{chart}(a). Figure \ref{SNRsim} illustrates the behaviour of the SNR as a function of pump pulse duration.  In those simulations the parametric gain is kept constant ($g=1.45$), while $\tau_{p}$ is increased up to 1000ps with the unbinned pixels [Fig.~\ref{SNRsim}(a)], up to 250ps with the 2x2 pixel binning [Fig.~\ref{SNRsim}(b)]; in such a way $\langle N_1\rangle$ (lower horizontal scale) increases almost linearly with the pump pulse duration (upper axis) up to 3500 photons per pixel and the SNR increases with the square root of $\langle N_1\rangle$ as predicted by Eqs.~(\ref{SNRclass}) and (\ref{SNRpdc}). 

The lower curves (black squares) represent the SQL obtained with the splitted coherent beams, while the upper curves (hollow triangles) reproduce the results obtained with the PDC source.
With the chosen parameters, the coherence length $x_{\rm FWHM}$ in the detection plane was evaluated on the order
of $10\mu$m. As discussed in Sec.\ref{sec5} (see Fig.~\ref{sigmavsxshift}), only the $40\mu$m pixels size allows to approach the optimal degree of correlation determined by the finite quantum efficiency of the detectors, i.e. $\sigma=1-\eta=0.1$. 
For this reason the factor of improvement $R$ obtained with the unbinned pixels [Fig.~\ref{SNRsim}(a)] is only $\sim 1.9$, while with the 2x2 pixel binning [Fig.~\ref{SNRsim}(b)] we are close to the optimal value $R=2.9$ (obtained from Eq.~(\ref{improvement}) with $\sigma=0.1$, $E_n\approx 0.5$). From these results it clearly emerges that the size of the coherence area determined by the optical setup imposes a lower limit to the size of the object details that can be resolved with a sensitivity beyond the SQL.

The dashed lines in the figure are obtained by using the expressions of the two-mode calculations (\ref{SNRpdc}) and (\ref{SNRclass}) and fit well the numerical results. The values of the relevant fitting parameter $\sigma$ are indicated beside each plots ($E_n$ is on the order of unity and does not affect the SNR value significantly, as can be inferred from expression (\ref{SNRpdc})). It should be noticed that both $R$ and $\sigma$ are almost independent on the pump pulse duration, as long as the parametric gain $g$ is fixed.  
   
\begin{figure}[ht]
\centering
\includegraphics[width=15.0cm]{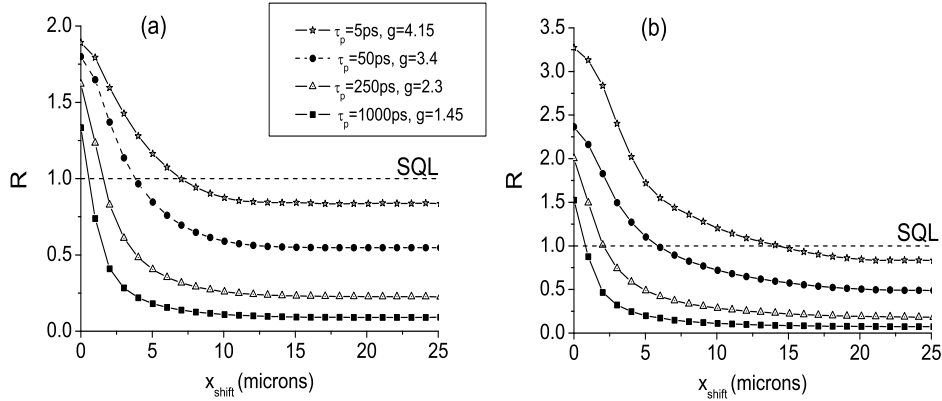}
\caption{The ratio $R=SNR_{\sigma}/SNR_{SQL}$ is plotted as a function of $x_{shift}$ for different values
of the pump pulse duration, considering $20\mu$m (a) and $40\mu$m (b) pixel detectors. All other parameters are the same as in Fig.~\ref{sigmavsxshift}.}
\label{Xshift}
\end{figure}

The images of Fig.~\ref{chart} allow to appreciate visually
the improvement that can be achieved with the PDC source (b) with respect to the classical source (c)
with the $40\mu$m pixel size. The  indicated SNR values are those obtained with a 125ps pump pulse (giving $\sim 1750$ photons/pixel) corresponding to the dotted vertical line in Fig.~\ref{SNRsim}. 

Finally we investigated on the sensitivity of the SNR improvement with respect to inaccuracies in the determination of the center of symmetry of the PDC far field.  In Fig.~\ref{Xshift} we plot the ratio $R=SNR_{\sigma}/SNR_{SQL}$ as a function of $x_{shift}$ for different values of the pump pulse duration and for a pixel detector size of $20\mu$m (a) and $40\mu$m (b). The chosen parameters are the same as in Fig.~\ref{sigmavsxshift}: the parametric gain $g$ is diminished for increasing values of $\tau_{p}$, so that the number of photons detected on the pixel area remains unchanged, about 3500 ph./pixel.

A first relevant issue is the deterioration of the correlation and the SNR that occurs at high parametric gains even for $x_{shift}=0$. This effect is related to the broadening of the coherence area which occurs at high parametric gains due to the finite transverse size of the pump pulse discussed in \ref{sec4b}. It becomes particularly relevant for the 5ps pump pulse with $g=4.15$ (black squares in the figures): the correlation factor $\sigma$ goes rather close to unity (see Fig.~\ref{sigmavsxshift}) and the SNR is at best close to the SQL value, for $x_{shift}=0$, as can be seen from Fig.~\ref{Xshift}.

Furthermore, for $\tau_{p}=5$ps the excess noise is extremely high ($E_n\sim 100$). As a consequence,
we see that the slightest inaccuracy in the position of the symmetry center leads to a
transition from sub-shot noise to strongly above shot noise signal/idler correlation (black square in Fig.~\ref{Xshift}).
For this reason the SNR rapidly decreases well below the SQL as $x_{shift}$ increases,
as can be seen from Fig.~\ref{Xshift}. 
The situation improves as the pulse duration is increased and the parametric gain is decreased, since
the excess noise diminishes and the SNR becomes more robust toward $x_{shift}$.
For example, considering the best detection conditions with the $40\mu$m pixel size [Fig.~\ref{Xshift}b], 
the tolerance on $x_{shift}$ (in order to have $R>1$) is less than $\sim 2\mu$m for $g=4.15$, while it increases to $\sim 14\mu$m with $g=1.45$.

To conclude, our simulations show that the detrimental effects of the broadening of the 
coherence area added to the fragility against imperfections in the measurement process become a much less relevant issue in the regime of low-gain, long-pulse regime.
 
\section{Conclusions}
We have proposed an imaging scheme that uses the multi-mode twin beams generated in the process of single-pass parametric down-conversion to detect the spatial distribution of faint objects through a differential measurement technique.  The object is located in the PDC far-field and the image is retrieved by subtracting pixel by pixel
the measured signal and idler photon fluxes. In this way the detection scheme can achieve a sensitivity beyond the SQL by exploiting the local character of the quantum correlations of the PDC far field, a property which was demonstrated experimentally in \citet{bib14,bib15} in a regime of high parametric gain. 

We developed a stochastic numerical model which includes realistic features of the imaging system in order to verify to which extent the PDC source allows to enhance the SNR of the measurement with respect to that obtained from a classical source. We pointed out the fragility of the sub-shot noise correlation 
against imperfections of the detection system in presence of a large amount of excess noise.
In particular we analyzed the effect of small errors (on the order of a few microns)
in the determination of the center of symmetry of the PDC far field, showing that they lead to a loss
of correlation proportional to the excess noise of the source. Other experimental imperfections, such as unbalances due to different losses in the test and reference arms and light scattering from the background, produce a similar deterioration of the sub-shot noise correlation. The quantum character of the correlation is therefore rapidly lost as the parametric gain (and therefore the excess noise) is increased. 

We suggest that the use of pump pulses with duration on the order of a few nanoseconds or more, much longer than the PDC coherence time, would allow to generate the same number of photons than in the case of the picoseconds pulses used in \cite{bib14} with a much lower parametric gain and a negligible excess noise.
Our simulations demonstrate that this represents a clear advantage in that the sub-shot noise correlation
becomes more robust and can be maintained up to gains corresponding to several thousands of PDC photons per pixel, a regime where the noise of the CCD can be neglected \cite{bib24}. 
We used the same numerical model to simulate the imaging experiment, showing that under these conditions 
the use of a PDC source lead to a substantial SNR enhancement with respect to classical
imaging in a domain of parameters accessible to experimental implementations.

Moreover, from our analysis it emerges that the degree of quantum correlations, and hence the improvement in the
SNR depends critically on the ratio between the pixel area and the far-field coherence area. The latter sets a lower limit on the size of the details that can be resolved with sensitivity above the SQL, and thus provides a resolution limit for high-sensitivity imaging with twin beams.

\section{Acknoledgements}
This work was carried out in the framework of the PRIN project of MIUR ``Twin beams in Quantum Imaging applications
and metrology''.

\noindent \vspace{1.cm}


\begin{thebibliography}{}
\bibitem{bib24}
Experimental results obtained in Como with a type I BBO crystal with a cw pump tend to confirm
the result of our numerical model and will be the object of a future pubblication.
\end{thebibliography}
\end{document}